\documentclass[twocolumn,twoside,slac_two,floatfix]{revtex4}
\pdfoutput=1
\usepackage{graphicx}
\usepackage{fancyhdr}
\begin{document}

\title{Analysis of the Impulsive Phase of Solar Flares with Pass 8 LAT data}

\author{R. Desiante}
\affiliation{Udine University, INFN-Trieste \& C.I.F.S., Italy}

\author{F. Longo}
\affiliation{Trieste University \& INFN-Trieste, Italy}
\author{N. Omodei}
\affiliation{Stanford University/KIPAC, Stanford, CA, USA}
\author{M. Pesce-Rollins}
\affiliation{INFN-Pisa, Italy}
\author{V. Pelassa}
\affiliation{University of Alabama in Huntsville/CSPAR, Huntsville, AL, USA}
\author{on behalf of the \textit{Fermi}-LAT Collaboration}

\begin{abstract}
We show the results of analyses performed on high-energy gamma-ray emission during the impulsive phase of solar flares detected by the LAT using Pass 8 data.  
We compare results obtained with Pass 7 and Pass 8 data sets, using both LAT Low Energy and standard data classes.  
With a dedicated event selection, Pass 8 allows standard analysis during the impulsive phase: it has been designed to be less susceptible to pile-up in the LAT Anti-Coincidence Detector caused by the intense hard X-ray emission at early times.
\end{abstract}

\maketitle

\thispagestyle{fancy}

\section{INTRODUCTION}
A solar flare is an intense and rapid energy release in the solar corona driven by stored magnetic energy liberated by coronal magnetic reconnection processes. This energy release results in acceleration of particles, including electrons, protons and heavy nuclei, to a wide range of energies and in heating of coronal and chromospheric plasma.

Looking at the X and gamma-ray light curves we can distinguish four different temporal phases of solar flares emission \cite{Hudson2011} \cite{Priest1981}:

- precursor, observed as a gradual raise of emission mainly visible in soft X-rays;

- impulsive, characterized by a rapid raise of hard X and gamma-ray flux;

- gradual, slow decaying of X and gamma-ray flux;

- extended, mainly observed as a sustained gamma-ray emission that can lasts for several hours.

Coronal Mass Ejections (CMEs) are also often observed in close association with gamma-ray detected solar flares.

\subsection{The impulsive phase of Solar Flares as seen by the LAT} \label{section_sf_LAT}

The Gamma-ray Burst Monitor (GBM) and the Large Area Telescope (LAT), the two instruments on-board to the \textit{Fermi} Observatory, can detect photons with energies from 8 keV up to 300 GeV. Both instruments also have very large fields of view (FOV) achieving together an unprecedented coverage of the X and gamma-ray sky: the GBM FOV consists of the whole not-occulted sky and the LAT scans about the 20\% of the sky at any instant.
These characteristics make the \textit{Fermi} Spacecraft a perfect observatory to study and monitor both the quiescent phase and the eruptive phases of solar activity at high energies
\cite{QuietSunPaper} \cite{2010FlarePaper} \cite{2012FlarePaper} \cite{AllafortPaper} \cite{behind-the-limb-flare_Omodei_Melissa}.

The first impulsive solar flare detected by \textit{Fermi} occurred on 2010 June 12 00:30 UT: together the \textit{Fermi} LAT and the GBM observed X and gamma-ray emission, from few keV up to $\sim$ 400 MeV, in coincidence with a moderate GOES M2.0 class solar flare. 
As fully explained in \cite{2010FlarePaper}, the observed spectrum has been interpreted as: 
\begin{itemize}
\item[-] electron bremsstrahlung, nuclear lines and pion decay components for energies $<$ 10 MeV;
\item[-] high-energy electron bremsstrahlung or pion decay component above 30 MeV.
\end{itemize}
The analysis of LAT data was performed using only the LAT Low Energy Events Technique (LLE) \cite{PelassaPaper} because the intense X-ray flux occurring during the prompt phase of a solar flare causes pile-up in the anti-coincidence detector (ACD) of the LAT \cite{2010FlarePaper} \cite{Abdo2009} resulting in a strong suppression of the rate of standard LAT Pass 6 / Pass 7 on-ground photon classification \cite{solitoAtwood2009}.
We show that these issues have been carefully addressed in new Pass 8 photon classification.

LLE event selection, that does not suffer of ACD pile-up, uses less discriminating criteria then the standard on-ground processing, resulting in a larger effective area but a lower signal-to-noise ratio: LLE data are background dominated and not suitable for localization studies.

\section{PASS 8 DATA: IMPROVEMENTS FOR SOLAR FLARES SCIENCE}

The event selection developed for the LAT has been periodically updated.
While Pass 7 REP is the current event analysis distributed to the community, the new Pass 8 data, that will be available in next few months, represent a radical revision of the entire event-level analysis that includes every aspect of the data reduction process. 
The improvements include a significant reduction in background contamination, an increased effective area, a better point-spread function, a better control on the systematic uncertainties and an extension of the energy range below 100 MeV and above a few hundred GeV \cite{ProcSymp2012} \cite{BaldiniTevpa}.  
This means to improve the solar flares detection capabilities of the LAT, 
in particular at low energies; the increase in photon statistics will also allow to better constrain the spectral features and to reduce the uncertainties in localization studies.
As already mentioned in Sec.\ref{section_sf_LAT}, a solar-flare dedicated event class selection has been also developed: this will alleviate the ACD pile-up effect often present during impulsive solar flares \cite{MelissaTevpa}.

\begin{figure*}[HTB!]
\centering
\includegraphics[scale=0.35]{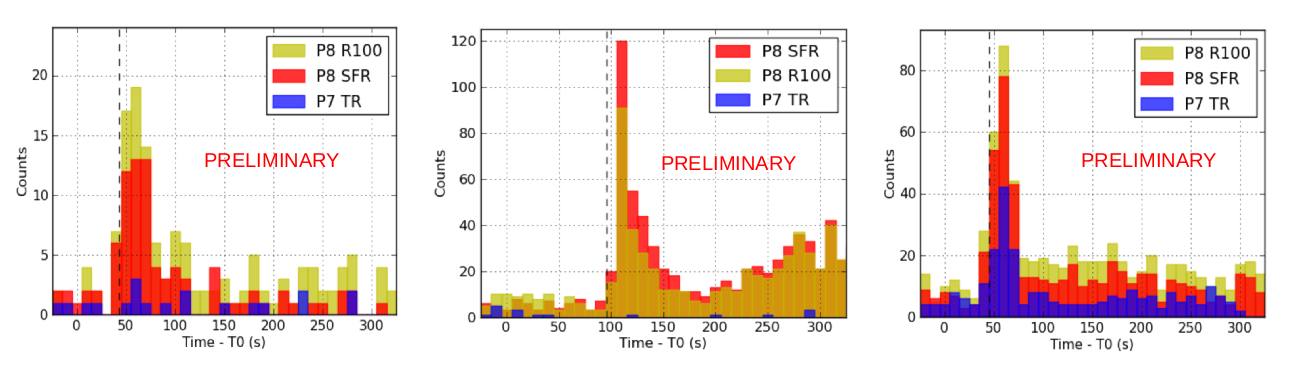}
\caption{SOL2010-06-12 (left panel), SOL2011-09-06 (middle panel), SOL2012-06-03 (right panel). 
For each solar flare we compare the light-curves obtained using different event selections. Data are extracted in the energy range 35 MeV - 10 GeV. T0 is the GBM trigger time; the dashed line marks the LLE detection time.} \label{all_lc}
\end{figure*}
\begin{figure*}[HTB!]
\centering
\includegraphics[scale=0.35]{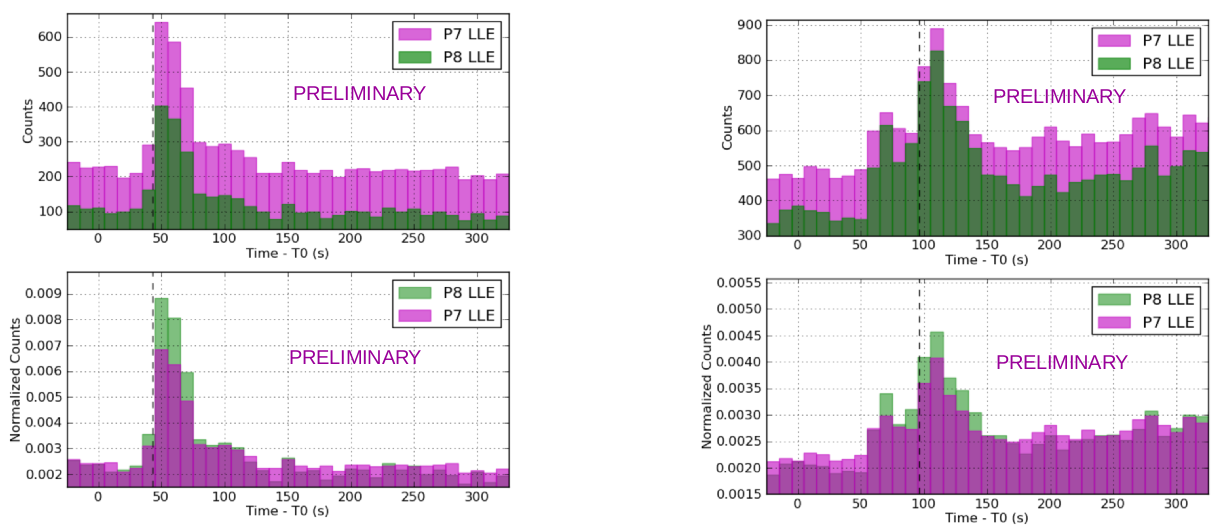}
\caption{Light-curves obtaind using Pass 8 LLE data VS P7 LLE data for SOL2010-06-12 (left panel) and SOL2011-09-06 (right panel). For each flare, the upper plot shows the number of detected counts while the bottom plot shows the number of detected counts normalized to the total.} \label{LLE_lc}
\end{figure*}
\begin{figure*}[hbt!]
\centering
\includegraphics[scale=0.35]{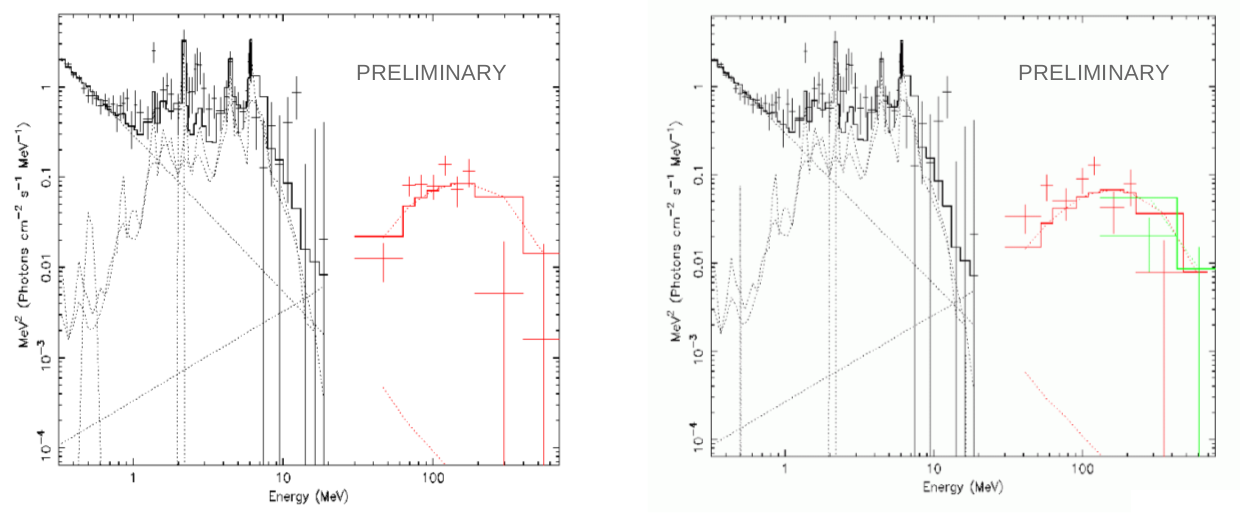}
\caption{SOL2011-09-06: background subctracted count spectra obtaind using GBM-BGO (black), P7 LLE (red, left panel), P8 LLE (red, right panel) and P8 R100 data (green, right panel).} \label{sp11}
\end{figure*}
\begin{figure*}[htb!]
\centering
\includegraphics[scale=0.35]{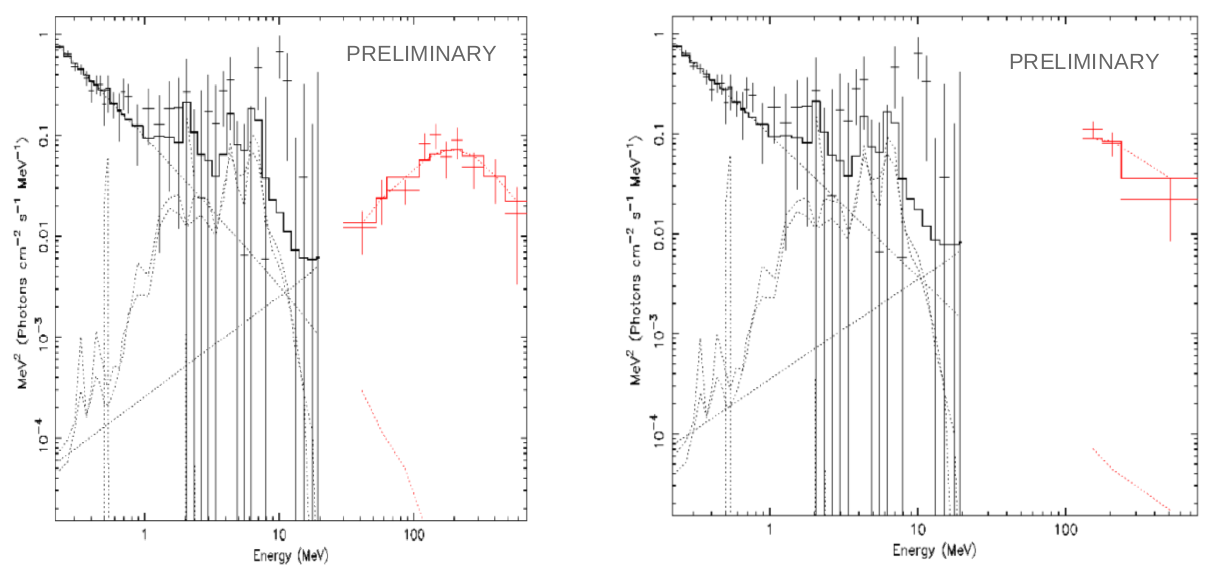}
\caption{SOL2012-06-03: background subctracted count spectra obtaind using GBM-BGO (black), P7 LLE (red, left panel) and P8 R100 data (red, right panel).} \label{sp12}
\end{figure*}

\subsection{Data analysis: Light-curves and Spectra}

Light-curves in Fig.\ref{all_lc} show a comparison of different LAT TRANSIENT data collected in the energy range 35 MeV - 10 GeV: 
\begin{itemize}
\item[-] P7 TR and P8 R100 are the loosest event classes for the the two different response functions Pass 7 REP and Pass 8;
\item[-] P8 SFR is the solar flare optimized event class newly developed in Pass 8.
\end{itemize}

We consider LAT observations of three different solar flares: 
\begin{itemize}
\item[-] SOL2010-06-12, M2.0 GOES class, already mentioned in Sec.\ref{section_sf_LAT};
\item[-] SOL2011-09-06, X2.1 GOES class, detected on 2011 September 06 22:17 UT;
\item[-] SOL2012-06-03, M3.3 GOES class, occurred on 2012 June 03 17:53 UT.
\end{itemize}

While impulsive and sustained gamma-ray emission from SOL2012-06-03 has been significantly detected using Pass 7 standard event classes, SOL2010-06-12 and SOL2011-09-06 were detected in Pass 7 only through LLE technique \cite{AllafortPaper}.

If we focus on standard event selections, Fig.\ref{all_lc} shows that Pass 8 performs better both on previously detected and not-detected flares. 
Moreover, the P8 SFR event class, developed with a better treatment of ACD variables, produces a less noisy light-curve for all flares. In the case of SOL2011-09-06 (the highest GOES class flare of our sample), this results also in a greater number of total counts detected; the Pass 7 signal is instead completely suppressed because of ACD pile-up caused by the intense X-ray flux.

We also tested the improvements of Pass 8 event selection on LLE technique. 
In Fig.\ref{LLE_lc} there is a comparison of light-curves obtained using Pass 8 LLE data versus Pass 7 LLE data for SOL2010-06-12 (left panel) and SOL2011-09-06 (right panel). 
The number of detected counts (upper plots) is higher for P7 LLE since Pass 8 event selection is less affected by background contamination but, as shown by the normalized number of detected counts (bottom plots), the P8 LLE light-curves provide a better signal-to-noise ratio.

To test the benefit of using Pass 8 data for spectral analysis we produced LAT (40 MeV - 1 GeV) and GBM-BGO (0.3 - 40 MeV) spectra accumulated during the LLE-detection time range for SOL2011-09-06 and SOL2012-06-03. Using the tool XSPEC \footnote{\url{http://heasarc.gsfc.nasa.gov/xanadu/xspec/}} we fit the data with the usual components (Fig.\ref{sp11} and Fig.\ref{sp12}): a power-law for electron bremsstrahlung in the BGO energy range, a nuclear lines template, Gaussian lines at 0.511 MeV and 2.223 MeV (related respectively to positron-electron annihilation and neutron capture) and a pion template in the high-energy part of the spectrum \cite{2010FlarePaper}.
While a rigorous spectral analysis is beyond the scope of this presentation, we want to stress that:
\begin{itemize}
\item[-] spectral analysis is now possible using P8 standard event classes also during the impulsive phase of solar flares;
\item[-] for both flares analyzed, R100 data cover a slightly wider energy range compared with P7 LLE data.
\end{itemize}

\section{PROSPECTS AND CONCLUSIONS}
Pass 8 data allow to study the Impulsive Phase of solar flares with standard LAT selections. Moreover Pass 8 event reconstruction improvements also impact the LLE selection technique.
A dedicated Pass 8 solar flare events class, less susceptible to ACD pile-up, is in development.

Preliminary results using Pass 8 data are in agreement with Pass 7 and but show greater signal-to-noise ratios and promising improvements for spectral analysis.
Validation of Pass 8 data at low energy ($<$ 100 MeV) is on-going in order to address energy dispersion issues. 
We plan a systematic study of GBM-BGO bright solar flares, useful to better understand the high-energy emission processes occuring in the solar corona.

Pass 8 improvements will allow to better study the low energy gamma-ray part of the spectrum and discriminate between hadronic and leptonic origin.
Temporal studies on the onset of high-energy emission are also on-going.

\bigskip 
\begin{acknowledgments}
The \textit{Fermi}-LAT Collaboration acknowledges support for LAT development, operation and data analysis from NASA and DOE (United States), CEA/Irfu and IN2P3/CNRS (France), ASI and INFN (Italy), MEXT, KEK, and JAXA (Japan), and the K.A.~Wallenberg Foundation, the Swedish Research Council and the National Space Board (Sweden). Science analysis support in the operations phase from INAF (Italy) and CNES (France) is also gratefully acknowledged.
\end{acknowledgments}

\bigskip 

\end{document}